\documentclass[prd,preprint,showpacs]{revtex4}
\usepackage{graphicx}
\usepackage{latexsym}
\usepackage{amsmath}
\newcommand{\bq}{\begin{equation}}
\newcommand{\eq}{\end{equation}}
\newcommand{\bqn}{\begin{eqnarray}}
\newcommand{\eqn}{\end{eqnarray}}

\newcommand{\lb}{\label}
\begin{document}
\title{Exact scaling solutions and fixed points for general scalar field}
\author{Yungui Gong}
\email{gongyg@cqupt.edu.cn}
\affiliation{College of Electronic Engineering, Chongqing
University of Posts and Telecommunications, Chongqing 400065,
China}
\affiliation{CASPER, Physics Department, Baylor University,
Waco, TX 76798, USA}
\author{Anzhong Wang}
\email{anzhong_wang@baylor.edu}
\affiliation{CASPER,
Physics Department, Baylor University, Waco, TX 76798, USA}
\author{Yuan-Zhong Zhang}
\affiliation{ CCAST (World Laboratory), P.O. Box 8730, Beijing 100080, China}
\affiliation{ Institute of Theoretical Physics, Chinese Academy of Sciences,
P.O. Box 2735, Beijing 100080, China}
\begin{abstract}
We show that the most general dark energy model that possesses a scaling solution
$\rho_\phi\propto a^n$ is the k-essence model, which includes both of the quintessence
and tachyon models. The exact scaling solutions are then derived. The potential
that gives the tracking solution in which dark energy exactly tracks the background
matter field is the inverse squared potential. The quintessence field with exponential
potential can be obtained from the k-essence field with the inverse squared potential.
We also find the fixed points and study their main properties, whereby
the scalar field dominant fixed point is identified.
\end{abstract}
\pacs{98.80.Cq}
\preprint{gr-qc/0603050}
\maketitle

\section{Introduction}

Recent cosmological observations of large-scale structure, type Ia
supernova and the cosmic microwave background anisotropy suggests
that the universe is currently experiencing an accelerated
expansion \cite{obs}. To produce such an acceleration in general
relativity, we have to introduce an ``unseen" component with a
large  negative pressure to the matter contents of the universe,
which is usually referred to as dark energy. This component accounts for
about seventy percents of the total matter in the universe. The simplest
candidate of dark energy is a very small cosmological constant.
Due to its unknown origin, many other dynamical dark energy models
have also been pursued, such as string/M-theory inspired models \cite{OH05} and brane-world
models \cite{Sa05}. For a review of dark energy models, please see \cite{DE}.

One class of models of particular interest is the dynamical models that have scaling attractor solutions.
It is well known that for the scalar field with the canonical kinetic term, exponential potentials have
scaling attractor solutions \cite{clw,ferreira}. For tachyon field, the potential $V(\phi)=V_0\phi^{-2}$
gives rise to scaling attractor solutions \cite{tachyon,lazkoz}. For a general scalar field that
interacts with the matter by a specific form, it was found that the general dark energy model
which gives rise to scaling attractor solutions has the Lagrangian density $L(X,\phi)=Xg(Xe^{\lambda\phi})$
\cite{tsujikawa}, where $X=\dot{\phi}^2/2$. After re-defining the scalar field, this Lagrangian density
is equivalent to the k-essence one $L(X,\phi) = V(\varphi)f(X)$ with $V(\varphi)=V_0\varphi^{-2}$.
This k-essence model gives the scaling attractor
solution when there is no interaction between the scalar field and the matter.

It is interesting to find the most general dark energy model that gives the scaling solution
$\rho_\phi\propto a^n$ when the specific interaction between the scalar field and the matter is absent.
Will the most general Lagrangian density still be $L(X,\phi)=Xg(Xe^{\lambda\phi})$ or
$L(X,\phi)=V(\varphi)f(X)$ with $V(\varphi)=V_0\varphi^{-2}$? The answer to this question is
not clear from the discussions in \cite{tsujikawa}. The question is addressed in this paper.
We find that the most general model that gives the scaling solution is the k-essence
one. To better understand the scaling solution, it is necessary to study the dynamical
properties of the k-essence solutions. In this paper, we are particularly interested in finding the potential for a
general scalar field with constant equation of state $w_\phi$ and studying
the properties of the fixed points of the solutions.

The paper is organized as follows.
In section II, we review the derivation of the general dark energy model $L(X,\phi)=Xg(Xe^{\lambda\phi})$
that gives rise to the scaling attractor solution with interaction and prove
that the k-essence model is the most general dark energy model that gives rise
to the scaling solutions with and without the interaction between the scalar field and the matter.
In section III,
we first review the exact
solutions for a canonical scalar field with constant $w_\phi$, then we derive
the exact solutions for the k-essence field with constant $w_\phi$. In section IV, we
discuss the properties of the fixed points for the potential $V(\phi)=V_0\phi^{-2}$. We
conclude our paper in section V.

\section{General dark energy model with tracking solutions}
For a general dark energy model with the Lagrangian density $L(X,\phi)=P(X,\phi)$, we get
the energy density $\rho_\phi=2XP_X-P$ and the pressure $p_\phi=P$ of the scalar field $\phi$,
where $P_X\equiv\partial P/\partial X$. By assuming a particular form of interaction between
the dark energy and the matter, we have the following Friedmann equations
\bq
\lb{sclreq2}
\dot{\rho_\phi}+3H(1+w_\phi)\rho_\phi=-Q\rho_m\dot{\phi},
\eq
\bq
\lb{matt1}
\dot{\rho}_m+3H(1+w_m)\rho_m=Q\rho_m\dot{\phi},
\eq
\bq
\lb{deneq}
\Omega_m+\Omega_\phi=1,
\eq
where $Q$ is assumed to be a constant, $\Omega_m=\rho_m/(3m^2_{pl} H^2)$,
$\Omega_\phi=\rho_\phi/(3m^2_{pl} H^2)$ and the reduced Planck mass $m^{-2}_{pl}=8\pi G$.
We are interested in the tracking solution in which the scalar field tracks the matter field so that
$\Omega_\phi/\Omega_m$ is a constant, therefore we require
\bq
\lb{cond1}
\Omega_m=1-\Omega_\phi={\rm constant},\quad
w_\phi={\rm constant}.
\eq
In terms of the variable $N=\ln a$, we get $\dot{\phi}=Hd\phi/dN$.
Eqs. (\ref{cond1}) and (\ref{deneq}) tell us that
\bq
\lb{cosmeq}
H^2=\frac{\rho_\phi}{3\Omega_\phi m^2_{pl}}=\frac{\rho_m}{3\Omega_m m^2_{pl}},
\eq
\bq
\lb{rel1}
\frac{d\ln\rho_\phi}{dN}=\frac{d\ln\rho_m}{dN}=\frac{d\ln H^2}{dN},
\eq
 and Eqs. (\ref{sclreq2}) and (\ref{matt1}) become
\begin{gather}
\lb{sclreq3}
\frac{d\rho_\phi}{dN}+3(1+w_\phi)\rho_\phi=-Q\rho_m\frac{d\phi}{dN},\\
\lb{mateq2}
\frac{d\rho_m}{dN}+3(1+w_m)\rho_m=Q\rho_m\frac{d\phi}{dN}.
\end{gather}
Substitute Eq. (\ref{rel1}) into Eqs. (\ref{sclreq3}) and (\ref{mateq2}), we get
\bq
\lb{rel2}
\frac{d\phi}{dN}=\frac{1}{Q}\left[\frac{d\ln\rho_m}{dN}+3(1+w_m)\right]
=-\frac{\Omega_\phi}{Q\Omega_m}\left[\frac{d\ln\rho_\phi}{dN}+3(1+w_\phi)\right]
=\frac{3\Omega_\phi}{Q}(w_m-w_\phi).
\eq
\bq
\lb{rel3}
\frac{d\ln\rho_\phi}{dN}=\frac{d\ln\rho_m}{dN}=\frac{d\ln H^2}{dN}=-3(1+w_{eff}),
\eq
where
\bq
w_{eff}=\Omega_m w_m+\Omega_\phi w_\phi.
\eq
From the definition $2X=H^2(d\phi/dN)^2$, we see that $X\propto H^2$. So we get
\bq
\lb{rel4}
\frac{d\ln X}{dN}=\frac{d\ln P}{dN}=\frac{d\ln p_\phi}{dN}=\frac{d\ln p_m}{dN}
=\frac{d\ln\rho_\phi}{dN}=-3(1+w_{eff}).
\eq
Since
$$\frac{d\ln P}{dN}=\frac{\partial \ln P}{\partial \ln X}\frac{d\ln X}{dN}+
\frac{\partial \ln P}{\partial \phi}\frac{d\phi}{dN},$$
so
\bq
\lb{rel7}
\frac{\partial \ln P}{\partial \ln X}-\frac{1}{\lambda}\frac{\partial \ln P}{\partial \phi}=1,
\quad \lambda=Q\frac{1+\Omega_m w_m+\Omega_\phi w_\phi}{\Omega_\phi(w_m-w_\phi)}.
\eq
Therefore, the general form of the Lagrangian is \cite{tsujikawa}
\bq
\lb{lag1}
P(X,\phi)=X g(Xe^{\lambda\phi}),
\eq
where $g(Y)$ is an arbitrary function.

From the above derivation, we see that $d\phi/dN$
is not a constant in general if the interaction is absent, $Q=0$. So when $Q=0$, the above
derivation is not applicable. However, the dark energy model (\ref{lag1}) gives tracking solutions even
when the interaction is absent. From Eq. (\ref{rel7}), we see that the tracking solutions give
$w_m=w_\phi$ if $Q=0$. In fact, if we make field transformation $\varphi=2\exp(\lambda\phi/2)/\lambda$,
then we get $X(\phi) g[X(\phi)e^{\lambda\phi}]=V(\varphi)F[X(\varphi)]$ with
$V(\varphi)=4/(\lambda^2\varphi^2)$ and $F(Y)=Yg(Y)$. So, the general dark energy model with
tracking solutions is actually a particular k-essence model.

From the definition, we get $w_\phi=P/(2XP_X-P)$. So a constant $w_\phi$ tells us that
$\partial \ln P/\partial \ln X=(1+w_\phi)/w_\phi$ is a constant. Therefore, we obtain
\bq
\lb{eq1}
\frac{\partial^2\ln P}{\partial \phi\partial \ln X}=0.
\eq
The general solution to Eq. (\ref{eq1}) is $P(X,\phi)=V(\phi)F(X)$,
which is the k-essence Lagrangian density. Therefore,  we conclude that the k-essence is
the general scalar field that gives rise to the scaling solution $w_\phi=$ constant.

\section{The K-essence Field}

In this section, we look for the potential of the k-essence field that gives the
scaling solutions. If we take $g(Y)=1-c/Y$, then we have
$Xg(Xe^{\lambda\phi})=X-c\exp(-\lambda\phi)$. Because the model $L(X,\phi) = Xg(Xe^{\lambda\phi})$
is equivalent to the k-essence model with the inverse squared potential, so the quintessence model
with the exponential potential is a particular case
of the k-essence with inverse squared potential.  Therefore,
we first review the exact solutions for the quintessence field with constant equation of state.

\subsection{Exact scaling solutions for quintessence field}

If $w_\phi$ is a constant, then the solution to Eq. (\ref{sclreq2}) with $Q=0$ is
\bq
\lb{rho1}
\rho_\phi=\frac{\dot{\phi}^2}{2}+V(\phi)=\rho_{\phi 0}\left(\frac{a_0}{a}\right)^{3(1+w_\phi)},
\eq
where the subscript $0$ means the current value of the variable.
From the definitions of the energy density and pressure, we find
\bq
\lb{phieq1}
\dot{\phi}^2
=\frac{2(1+w_\phi)}{1-w_\phi}V(\phi)=(1+w_\phi)\rho_\phi.
\eq
So Eq. (\ref{deneq}) becomes
\bq
\lb{freq}
H^2=\frac{1}{3m^2_{pl}}\left[\rho_{m0}\left(\frac{a_0}{a}\right)^{3(1+w_m)}+
\rho_{\phi 0}\left(\frac{a_0}{a}\right)^{3(1+w_\phi)}\right].
\eq
Take $dt=a^{3(1+w_\phi)/2}d\tau$, we get
\bq
\lb{atau}
\sqrt{\Omega_{\phi 0}}H_0 d\tau=\frac{da}{a\sqrt{1+\Omega_{m0}/\Omega_{\phi 0} a^{3(w_\phi-w_m)}}}.
\eq
Thus the solutions are \cite{agff}
\bq
\lb{atausol}
a(\tau)=\begin{cases}
\left(\frac{\Omega_{m0}}{\Omega_{\phi 0}}\right)^{1/3(w_m-w_\phi)}\left[\sinh\left(
\frac{3(w_m-w_\phi)}{2}\sqrt{\Omega_{\phi 0}}H_0\tau\right)\right]^{2/3(w_m-w_\phi)}&
\text{for $w_m\neq w_\phi$},\\
\exp[H_0(\tau-\tau_0)]& \text{for $w_m=w_\phi$}.
\end{cases}
\eq

Since $\dot{\phi}=aH d\phi/da$, taking $a_0=1$ and combining Eqs. (\ref{phieq1}) with (\ref{freq}), we get
\bq
\lb{phia}
\left(\frac{d\phi}{da}\right)^2=
\frac{3m^2_{pl}(1+w_\phi)\Omega_{\phi 0}}{\Omega_{m0}a^{2+3(w_\phi-w_m)}+\Omega_{\phi 0}a^2}.
\eq
The solution to the above equation is
\bq
\lb{phia1}
\phi(a)=\begin{cases}
\pm\frac{2\sqrt{3m^2_{pl}(1+w_\phi)}}{3(w_m-w_\phi)}{\rm Arsh}\left(
\sqrt{\frac{\Omega_{\phi 0}}{\Omega_{m0}}}a^{3(w_m-w_\phi)/2}\right)+\phi_{in} &
\text{$w_m\neq w_\phi$ and $\Omega_{m0}\neq 0$},\\
\pm \sqrt{3m^2_{pl}(1+w_\phi)\Omega_{\phi 0}}\ln a +\phi_{in}&
\text{$w_m=w_\phi$ or $\Omega_{m0}=0$},
\end{cases}
\eq
where $\phi_in$ is an arbitrary constant determined by the initial condition.
One interesting case is when $\Omega_{\phi 0}\ll \Omega_{m0}$ and $w_m\neq w_\phi$. In this case, we have
\bq
\lb{phia2}
\phi(a)=\pm \frac{2}{3(w_\phi-w_m)}\sqrt{\frac{3m^2_{pl}(1+w_\phi)\Omega_{\phi 0}}{\Omega_{m0}}}
a^{-3(w_\phi-w_m)/2}+\phi_{in}, \quad {\rm if}\ \Omega_{\phi 0}\ll \Omega_{m0}.
\eq
Then, combining Eqs. (\ref{rho1}), (\ref{phieq1}) and (\ref{phia1}),
we obtain the potential of the scalar field \cite{poten}
\bq
\lb{vphi}
V(\phi)=\begin{cases}
\frac{1-w_\phi}{2}\rho_{\phi 0}\left[\sqrt{\frac{\Omega_{m 0}}{\Omega_{\phi 0}}}
\sinh\left(\frac{3(w_m-w_\phi)}{2\sqrt{3(1+w_\phi)}}\frac{\phi-\phi_{in}}{m_{pl}}
\right)\right]^{-2(1+w_\phi)/(w_m-w_\phi)} &
\text{$w_m\neq w_\phi$ and $\Omega_{m0}\neq 0$},\\
\frac{1-w_\phi}{2}\rho_{\phi 0}\exp\left( -\sqrt{\frac{3(1+w_\phi)}{\Omega_{\phi 0}}}
\frac{\phi-\phi_{in}}{m_{pl}}\right)&
\text{$w_m=w_\phi$ or $\Omega_{m0}=0$},\\
\frac{1-w_\phi}{2}\rho_{\phi 0}\left[\frac{3(w_\phi-w_m)}{2}
\sqrt{\frac{\Omega_{m0}}{3(1+w_\phi)\Omega_{\phi 0}}}
\frac{\phi-\phi_{in}}{m_{pl}}\right]^{2(1+w_\phi)/(w_\phi-w_m)}&
\text{$\Omega_{\phi 0}\ll \Omega_{m0}$ and $w_m\neq w_\phi$}.
\end{cases}
\eq

\subsection{Exact scaling solutions for the k-essence field}

The Lagrangian density for the k-essence field is
$$L=-V(\phi)F(X).$$
From the above Lagrangian density, we find
\bq
\lb{kden}
\rho_\phi=V(\phi)[F-2XF_X],\quad p_\phi=-V(\phi)F(X),
\eq
where $F_X\equiv dF/dX$. Then, from the definition of $w_\phi$ we find
\bq
\lb{rel5}
\frac{d\ln F(X)}{d\ln X}=\frac{X_cF_X(X_c)}{F(X_c)}=\frac{1+w_\phi}{2w_\phi}.
\eq
So, $X_c$ is a constant along the scaling solution and its value is
determined by   Eq. (\ref{rel5}). Using this fact, we obtain
\begin{gather}
\lb{dphida1}
\frac{d\phi}{da}=\pm \frac{\sqrt{2X_c}}{Ha}=\pm \frac{\sqrt{2X_c}a^{3(1+w_\phi)/2-1}}{H_0
\sqrt{\Omega_{\phi 0}+\Omega_{m0}a^{3(w_\phi-w_m)}}},\\
\lb{tachp1}
V(\phi)=-3m^2_{pl}\Omega_{\phi 0} H^2_0\frac{w_\phi}{F(X_c)}a^{-3(1+w_\phi)}.
\end{gather}

(1) $w_m=w_\phi$ or $\Omega_m=0$. In this case, we have
\begin{gather}
\lb{aphi2}
a(\phi)=\left[\pm \frac{3(1+w_\phi)}{2\sqrt{2X_c}}H_0(\phi-\phi_{in})\right]^{2/3(1+w_\phi)},\\
\lb{vphi2}
V(\phi)=-\frac{8\Omega_{\phi 0}w_\phi X_c}{3(1+w_\phi)^2 F(X_c)}\left(\frac{m_{pl}}{\phi-\phi_{in}}\right)^2.
\end{gather}

(2) $w_m\neq w_\phi$ and $\Omega_m\neq 0$. In this case, letting $d\phi=a^{3(1+w_\phi)/2}d\varphi$, we find
\bq
\lb{aphi3}
a(\varphi)=\left(\frac{\Omega_{m0}}{\Omega_{\phi 0}}\right)^{1/3(w_m-w_\phi)}
\left[\sinh\left(\frac{3(w_m-w_\phi)}{2}\sqrt{\frac{\Omega_{\phi 0}}{2X}}H_0(\varphi-\varphi_{in})
\right)\right]^{2/3(w_m-w_\phi)},
\eq
and the potential is
\bq
\lb{vphi3}
\begin{split}
V(\varphi)=&-3m^2_{pl}H^2_0\Omega_{\phi 0}\frac{w_\phi}{F(X)}
\left(\frac{\Omega_{\phi 0}}{\Omega_{m0}}\right)^{(1+w_\phi)/(w_m-w_\phi)}\times\\
&\left[\sinh\left(\frac{3(w_m-w_\phi)}{2}\sqrt{\frac{\Omega_{\phi 0}}{2X}}H_0(\varphi-\varphi_{in})
\right)\right]^{2(1+w_\phi)/(w_\phi-w_m)}.
\end{split}
\eq
The solution for the scale factor is also given by Eq. (\ref{atausol}).

\section{Fixed points of tracking solutions}

In the previous sections, we show that the k-essence scalar field with the inverse
squared potential $V(\phi)\propto \phi^{-2}$ gives the most general tracking solutions
in which $w_\phi=w_m=$ constant. In this section, we discuss the fixed points for the
general tracking solutions. Without loss of generality, we write the k-essence potential
as $V(\phi)=4/(\lambda^2\phi^2)$, and set
$$x=\dot{\phi}=\sqrt{2X},\quad y=\frac{V(\phi)}{3m^2_{pl}H^2}=\frac{4}{3\lambda^2 m^2_{pl}H^2\phi^2},
\quad z=\frac{\rho_m}{3m^2_{pl}H^2}.$$
Then, the energy conservation equation for the scalar field $\dot{\rho}_\phi+3H(1+w_\phi)\rho_\phi=0$
gives
\bq
\lb{dxdn}
\frac{dx}{d\ln a}=-\frac{\sqrt{3}\lambda m_{pl}(F-x^2F_X)y^{1/2}+3xF_X}{F_X+x^2F_{XX}},
\eq
while the Friedmann Eq. (\ref{freq}) yields
\bq
\lb{dydn}
\frac{dy}{d\ln a}=-xy\left[\sqrt{3}\lambda m_{pl}y^{1/2}+\frac{3xF_X}{F-x^2F_X}\right]
+3yz\left[1+w_m+\frac{x^2F_X}{F-x^2F_X}\right],
\eq
and the energy conservation equation for the matter field $\dot{\rho}_m+3H(1+w_m)\rho_m=0$
together with Eq. (\ref{freq}) give
\bq
\lb{dzdn}
\frac{dz}{d\ln a}=3z(z-1)\left(1+w_m+\frac{x^2F_X}{F-x^2F_X}\right).
\eq
Eq. (\ref{freq}) gives the constraint equation
\bq
\lb{cond7}
y(F-x^2F_X)+z=1.
\eq
For the tachyon, $F(X)=\sqrt{1-2X}$, the above equations reduce to Eqs. (11-13) in Ref. \cite{lazkoz}.

The fixed points are those points that satisfy $dx=dy=dz=0$ in Eqs. (\ref{dxdn})-(\ref{cond7}).
So there are three fixed points:
(i) $x=y=0$ and $z=1$ which gives $\Omega_{\phi}=0$.
(ii) $z=0$ and $\sqrt{3}\lambda m_{pl}(F-x^2F_X)y^{1/2}=-3xF_X$. This is the scalar field dominated
case $\Omega_{\phi}=1$ and $\lambda^2=-3(1+w_\phi)F_X/m^2_{pl}$.
(iii) $\sqrt{3}\lambda m_{pl}(F-x^2F_X)y^{1/2}=-3xF_X$ and  $1+w_m+x^2F_X/(F-x^2F_X)=0$. This
is the case when the scalar field tracks the background field with $w_m=w_\phi$ and
$\Omega_\phi=-3(1+w_\phi)F_X/(m_{pl}\lambda)^2$.

Now let us discuss the stability of these critical points. We first consider the property of
the fixed point $x_c=y_c=0$. Let $x=x_c+\delta x$ and $y=y_c+\delta y$,
then Eqs. (\ref{dxdn}), (\ref{dydn}) and (\ref{cond7}) give the following equations for small perturbations
\begin{gather}
\lb{dxpert1}
\frac{d\delta x}{d\ln a}=-3\delta x-\frac{\sqrt{3}}{2}\lambda m_{pl}\frac{F}{F_X}y_c^{-1/2}\delta y,\\
\lb{dypert1}
\frac{d\delta y}{d\ln a}=3(1+w_m)\delta y.
\end{gather}
So the eigenvalues are $-3$ and $3(1+w_m)$. When $w_m>-1$, the fixed point is an unstable saddle point,
and when $w_m=-1$ the fixed point is a stable node.

For the fixed points (ii) and (iii),   Eqs. (\ref{dxdn}), (\ref{dydn}) and (\ref{cond7}) give us
the following equations for small perturbations
\begin{gather}
\lb{dxpert}
\frac{d\delta x}{d\ln a}=3w_\phi\delta x+\frac{3}{2}c_s^2 x_c y_c^{-1}\delta y,\\
\lb{dypert}
\frac{d\delta y}{d\ln a}=-3(1+w_\phi)x_c^{-1}y_c\left[\frac{w_m\Omega_\phi}{c_s^2}+1-\Omega_\phi\right]\delta x
-\left[3(w_m-w_\phi)\Omega_\phi+\frac{3}{2}(1+w_\phi)\right]\delta y,
\end{gather}
where the sound speed is defined as
\bq
\lb{cs2}
c_s^2=\frac{dp_\phi}{d\rho_\phi}=\frac{F_X(X)}{F_X(X)+x^2_cF_{XX}(X)}\biggr|_{X=x^2_c/2},
\eq
$x_c$ satisfies
\bq
\lb{wphi}
w_\phi=-\frac{F(X)}{F(X)-x^2_cF_X(X)}\biggr|_{X=x^2_c/2},
\eq
and
\bq
\lb{omphi}
\Omega_\phi=y_c[F(X)-x^2_cF_X(X)]|_{X=x^2_c/2}.
\eq
We need to find out the eigenvalues of the matrix
\bq
\lb{eigen}
M=\begin{bmatrix}
3w_\phi& 3c_s^2 x_c y_c^{-1}/2\\
-3(1+w_\phi)x_c^{-1}y_c\left[w_m\Omega_\phi/c_s^2+1-\Omega_\phi\right]&
\quad-3[(w_m-w_\phi)\Omega_\phi+(1+w_\phi)/2]
\end{bmatrix}.
\eq

For the fixed point (ii), we have $3^{1/2}\lambda m_{pl}y^{1/2}_c=-3x_c F_X/(F-x^2_c F_X)$,
$X=x^2_c/2$,
and $y_c=1/(F-x^2_c F_X)$. The eigenvalues are $3(w_\phi-w_m)$ and
$3(-1+w_\phi)/2<0$. So, the fixed point is a stable node for $w_\phi<w_m$ and an unstable
saddle for $w_\phi\ge w_m$.

For the fixed point (iii), we have $3^{1/2}\lambda m_{pl}y^{1/2}_c=-3x_c F_X/(F-x^2_c F_X)$
and $w_\phi=w_m$ is related with $x_c$ through Eq. (\ref{wphi}). The eigenvalues are
\bq
\lb{eingv1}
\frac{-3(1-w_m)\pm 3[(1+3w_m)^2-8(1+w_m)(w_m\Omega_\phi+(1-\Omega_\phi)c^2_s)]^{1/2}}{4},
\eq
where $c_s^2$ is given by Eq. (\ref{cs2}). Thus, the fixed point is a stable node if
$c_s^2>w_m$ \cite{kessence} and an unstable saddle point otherwise. For the tachyon field, the above
results reduce to those obtained in \cite{lazkoz}. All these results are summarized in table \ref{table1}.
\begin{table}[htp]
\caption{Summary of the properties of the critical points.}
\label{table1}
\begin{center}
\begin{tabular}{|c|c|c|c|c|}
  \hline
$x$&$y$&Stability&$\Omega_\phi$&$w_\phi$\\
  \hline
0&0&Unstable saddle for $w_m>-1$&0&Undefined\\\hline
$x_{c2}$ & $y_{c2}$ & Stable node for $w_\phi<w_m$ & 1 & Eq.(\ref{wphi}) \\
&&Unstable saddle for $w_\phi\ge w_m$&&\\\hline
$x_{c3}$&$y_{c3}$&Stable node for $c_s^2>w_m$&Eq. (\ref{omphi})&$w_m$\\
&&Unstable saddle for $c_s^2\le w_m$&&\\\hline
\end{tabular}
\end{center}
\end{table}

\section{Conclusion}

In \cite{tsujikawa}, it was found that when a specific interaction between the dark energy
and the matter is present, the most general dark energy
model that gives the tracking solution is the model $L(X,\phi)=Xg(Xe^{\lambda\phi})$.
In this paper, we prove that the k-essence models with the potentials given by Eqs. (\ref{vphi2})
and (\ref{vphi3}) are the most general dark energy models which possess a constant
equation of state. This result is applicable to the cases both with and without interactions
between the dark energy and the matter. Moreover, it also includes a variety of
interesting dark energy models, such as quintessence and tachyon.
The potential that gives rise to the tracking solution with $w_m=w_\phi$ is the inverse squared potential
$V(\phi)\propto \phi^{-2}$. This potential reduces to the exponential potential of the quintessence if we
choose $F(X)=X-c$. We also show that for a scalar field with constant equation of state,
there are two fixed points
relevant to dark energy: (a) the scalar field dominant attractor $\Omega_\phi=1$. (b) the tracking attractor
$w_m=w_\phi$. The properties of the fixed points are summarized in table \ref{table1}.
For the first fixed point, it is a stable node if $w_\phi<w_m$ and an unstable saddle if $w_\phi\ge w_m$.
For the dark energy model, we require $w_\phi<0$ and $w_m\ge 0$, so it is an attractor. For the second
fixed point, it is a stable node if $c_s^2>w_m$ and an unstable saddle if $c_s^2\le w_m$. Considering the late
time behavior of the dark energy, we set $w_m=0$. We also require the sound speed to be positive, so again the
fixed point is an attractor for relevant dark energy model.


\begin{acknowledgments}

Y. Gong was supported by Baylor University, NNSFC under grant No. 10447008,
SRF for ROCS, State Education Ministry
and CQUPT under grant No. A2004-05. Y.Z.
Zhang's work was in part supported by NNSFC under Grant No.
90403032 and also by National Basic Research Program of China
under Grant No. 2003CB716300.
\end{acknowledgments}

\end{document}